\documentclass[aps, superscriptaddress, twocolumn,prl,floatfix]{revtex4}
\usepackage{amssymb}
\usepackage{amsmath}
\usepackage{graphicx}
\usepackage{color}
\usepackage{natbib}
\usepackage{multirow}
\usepackage{upgreek}
\usepackage{capt-of}
\usepackage{pbox}

\def\Gq{e^2/h}

\def\Vtg{V_{TG}}
\def\Vbg{V_{BG}}

\def\nutg{\nu_{T}}
\def\nubg{\nu_{B}}

\begin{document}
\title{Selective equilibration of spin and valley polarized quantum Hall edge states in graphene}
\author{F. Amet}
\affiliation{Department of Applied Physics, Stanford University, Stanford, CA 94305, USA}
\author{J. R. Williams}
\affiliation{Department of Physics, Stanford University, Stanford, CA 94305, USA}
\author{K.Watanabe}
\affiliation{Advanced Materials Laboratory, National Institute for Materials Science, 1-1 Namiki, Tsukuba, 305-0044, Japan}
\author{T.Taniguchi}
\affiliation{Advanced Materials Laboratory, National Institute for Materials Science, 1-1 Namiki, Tsukuba, 305-0044, Japan}
\author{D. Goldhaber-Gordon}
\affiliation{Department of Physics, Stanford University, Stanford, CA 94305, USA}

\begin{abstract}
We report on transport measurements of dual-gated, single-layer graphene devices in the quantum Hall regime, allowing for independent control of the filling factors in adjoining regions. Progress in device quality allows us to study scattering between edge states when the four-fold degeneracy of the Landau level is lifted by electron correlations, causing edge states to be spin and/or valley polarized. In this new regime, we observe a dramatic departure from the equilibration seen in more disordered devices: edge states with opposite spins propagate without mixing. As a result, the degree of equilibration inferred from transport can reveal the spin polarization of the ground state at each filling factor. In particular, the first Landau level is shown to be spin-polarized at half-filling, providing an independent confirmation of a conclusion of Ref.~\cite{Young2012}. The conductance in the bipolar regime is strongly suppressed, indicating that co-propagating edge states, even with the same spin, do not equilibrate along PN interfaces. We attribute this behavior to the formation of an insulating $\nu$$\,=\,$0 stripe at the PN interface.
\end{abstract}
\maketitle

At low magnetic fields, the electronic properties of graphene are well described by a non-interacting Dirac Hamiltonian~\cite{Geim2007, Neto2009}, with four-fold degeneracy associated with spin and valley isospin, an additional degree of freedom due the hexagonal crystal lattice of graphene. As a result, graphene exhibits an anomalous quantum Hall effect with a transverse conductance quantized as 4(n+$\frac{1}{2}$)$\Gq$, where n is an integer~\cite{Zhang2005,Novoselov2005,Gusynin2005}.  At higher fields, Zeeman coupling and electron correlations can lift the four-fold degeneracy of the energy spectrum, resulting in spin- and valley-polarized Landau levels~\cite{Zhang2006}.  The nature of the ground state for each Landau level at partial filling depends on which symmetry-breaking energy dominates, a controversial topic over the years~\cite{Zhang2006, Jiang2007, Goerbig2011, Kharitonov2012, Hou2010, Barlas2012, Alicea2007, Alicea2007bis, Sheng2007, Checkelsky2008, Nomura2006, Jung2009, Goerbig2006, Yang2006}. While progress has been made in our understanding of the symmetry-breaking of partially filled Landau levels, direct observation of their polarization remains difficult. Previous work focused on the in-plane and perpendicular field dependence of the bulk 2D quantum Hall gaps~\cite{Young2012, Jiang2007, Checkelsky2008}, but an alternative approach would be to directly study edge transport.  

\begin{figure}[t!]
\center \label{fig1}
\includegraphics[width=3 in]{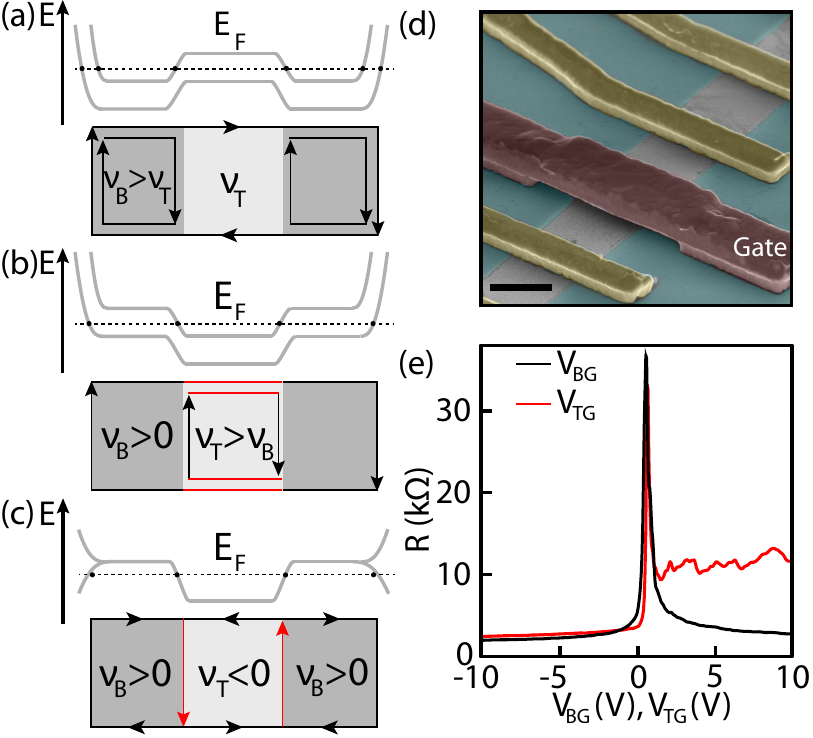}
\caption{\footnotesize{(a) Spatial variations of Landau levels in the case $\nubg>\nutg>0$: levels are bent by the confinement potential close to the edges of the device. Only $\nutg$ edge states are fully transmitted through the barrier. (b) Landau levels in the case $\nutg>\nubg>0$. Equilibration (shown in red) occurs along the physical edges of the flake under the top gate. (c) Landau levels in the case $\nutg\nubg<0$, with a degenerate zeroth level. This degeneracy is lifted close to the edges. Equilibration occurs along the PN interface. (d) False-colored electron micrograph of a top-gated device: the top-gate (red) is suspended above the graphene flake (gray). The whole device rests on a boron-nitride substrate (blue). The scale bar is 1$\mu$m. (e) Resistance $R$ as a function of $\Vbg$ and $\Vtg$, measured at a temperature $T$$\,$=2\,K.}}
\end{figure}

To this end, we measured the conductance of eight dual-gated graphene devices in the quantum Hall regime, at temperature T$\,$=$\,$250$\,$mK and for magnetic field B up to 14$\,$T. Our samples use a hexagonal boron-nitride substrate (h-BN), which greatly improves the electronic performance of graphene devices~\cite{Dean2009}, and a suspended top-gate~\cite{Amet2013, Velasco2012,Weitz2010}. The resulting quality of our devices allows us to study bipolar transport where the spin and valley four-fold degeneracies of the Landau levels are fully lifted.  When the filling factors under and outside the top-gate differ, the two-terminal conductance of such devices strongly depends on scattering between edge states, with new plateaus resulting from their mixing. The values of these plateaus suggest that edge states with different spin polarization do not equilibrate at the scale of our devices. This contrasts with the valley polarization, since intervalley scattering along the disordered edges of these samples causes edge states with different valley polarization --- but same spin --- to equilibrate. The pattern of equilibration for each pair of filling factors depends on the ground states at quarter- and half filling for the zeroth and first Landau levels. In particular, our measured conductance plateaus at filling factor $\nu$$\,=\,$4 are consistent with a spin-polarized first Landau level at half-filling. The conductance in the bipolar regime becomes vanishingly small as B increases, contrary to previous observations~\cite{Williams2007, Ozyilmaz2007,Velasco2012}, suggesting the formation of a narrow $\nu$$\,=\,$0 insulating stripe along the PN interface. 

We present here a unified data set from a single device, called Device A; additional data from similar devices are available in Ref.$\,$\cite{Suppinfo}. Device A is a 3$\,\upmu$m long, 1$\,\upmu$m wide graphene stripe, with a metallic top-gate suspended $\sim\,$90$\,$nm above the middle third of the device [Fig. 1(d)].  Details of the fabrication are described in Ref.~\cite{Amet2013,Suppinfo}. The two-terminal conductance $g$ is measured in a $^{3}$He cryostat using a conventional lock-in setup with a 100$\,$$\mu$V voltage-bias excitation at 137$\,$Hz. Resistance $R \equiv 1/g$ at B$\,=\,$0 as a function of the back-gate voltage $\Vbg$ shows residual doping $\delta$n $\sim$ 10$^{11}\,$cm$^{-2}$, peak resistance 30$\,$k$\Omega$ and mobility 120 000$\,$cm$^{2}$/Vs at T$\,$=$\,2\,$K [Fig. 1(e)], demonstrating the high quality of this sample. $R(\Vtg)$ shows the usual electron-hole asymmetry characteristic of graphene PN junctions. The back-gate capacitance extracted from magneto-transport measurements is 5.9x10$^{10}\,$cm$^{-2}$/V, and the top-to-back-gate capacitance ratio is 1.05, in good agreement with the geometry of the device as described in Ref.~\cite{Suppinfo}.

In the quantum Hall regime, the conductance depends on the filling factors under and outside the top-gated region, $\nutg$ and $\nubg$ respectively. When $\nutg$ is reduced below $\nubg$, only $\nutg$ edge states are fully transmitted through the top-gated region [Fig. 1(a)], while the others are fully reflected, allowing control of the edge states' trajectories. Two other regimes are of particular interest because the overall two-terminal conductance $g$ strongly depends on the interactions between edge states~\cite{Abanin2007}. In the unipolar regime, if $\vert\nutg\vert$$\,>\,$$\vert\nubg\vert$ the conductance depends on scattering between edge states under the gate. When Landau levels are degenerate, the conductance has been observed to be reduced from $\nubg\Gq$ and quantized as $g_{\nutg,\nubg}\,=\,\nutg\nubg$/(2$\nutg{-}\nubg$)~\cite{Williams2007,Ozyilmaz2007,Velasco2012}. This was attributed in Ref~\cite{Abanin2007} to the complete mixing of edge states: charge carriers are randomly scattered between edge states and have a probability 1/$\nutg$ to be ejected in any given edge state after propagating under the gate, regardless of the state they were injected into at the contacts~\cite{Abanin2007}. When the spin and valley degeneracies are lifted, however, it is not clear whether this model still applies since scattering between edge states might depend on their spin and/or valley polarization. In the bipolar regime, the role of scattering is even starker: edge states in the electron- and hole-doped regions have opposite chirality, so $g$ is nonzero only if these equilibrate while co-propagating along the PN interfaces [Fig. 1(c)]. Previous work showed that for dual-gated devices on an SiO$_{2}$ substrate, scattering along the PN interface is strong enough to completely mix edge states in this regime as well, resulting in new conductance plateaus at fractional multiples of the quantum of conductance $\Gq$ ~\cite{Williams2007, Ozyilmaz2007,Velasco2012}. 

\begin{figure}[t!]
\center \label{fig3}
\includegraphics[width=3 in]{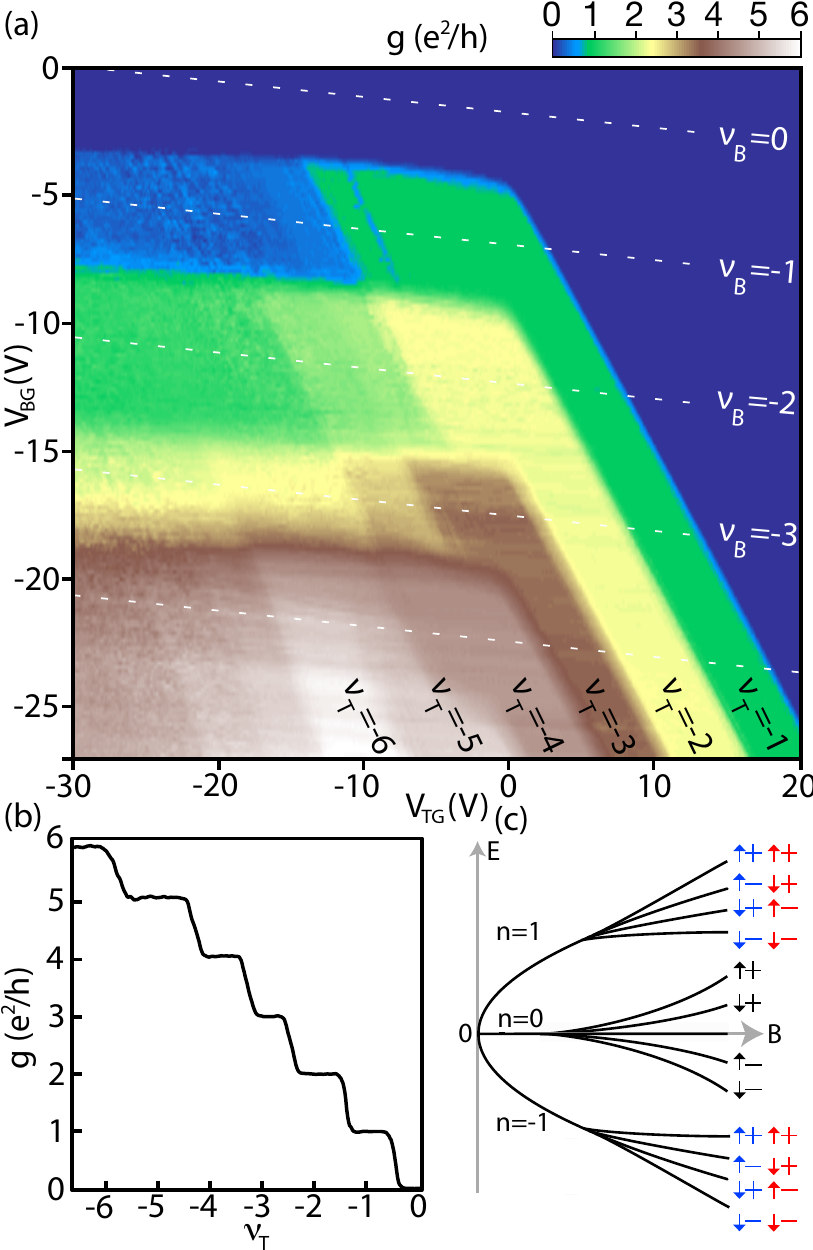}
\caption{\footnotesize{(a) Two-terminal conductance $g$ as a function of both gate voltages $\Vtg$ and $\Vbg$, measured at 14T, T$\,$=$\,$1K. (b) Partial transmission of edge states at $\nubg$$\,$=$\,$-6 as the filling factor under the top-gate is depleted. (c) Energy diagram for spin- (blue) or valley-polarized (red) first Landau level at half-filling.}}
\end{figure}

\begin{figure*}
\center \label{fig4}
\includegraphics[width=7 in]{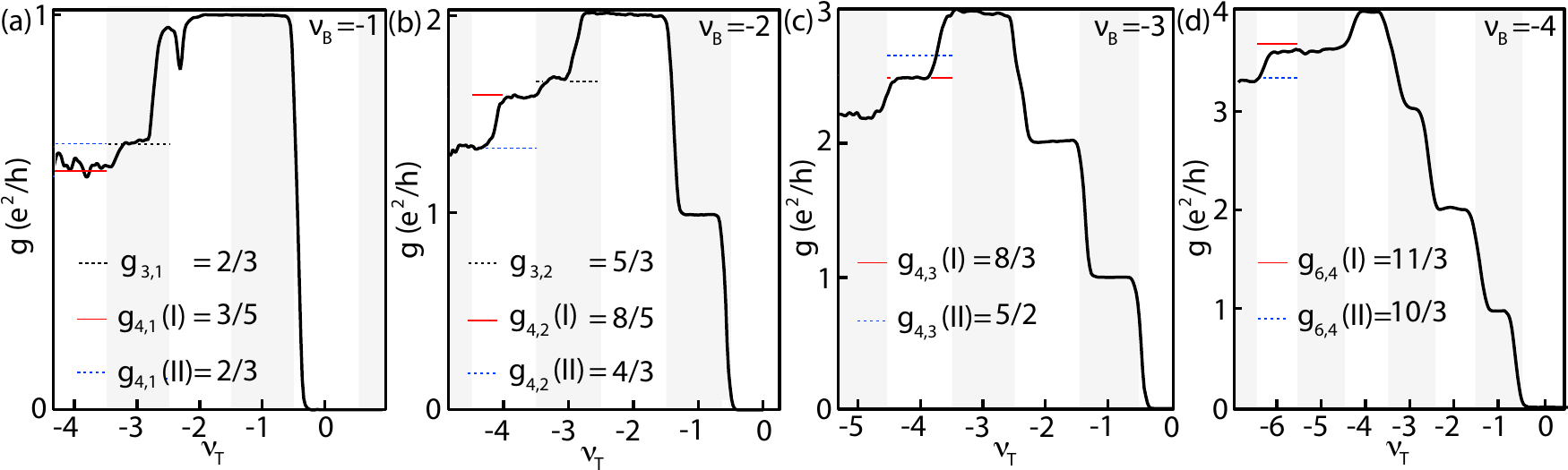}
\caption{\footnotesize{Cuts of the two-terminal conductance $g(\nutg)$ at constant $\nubg$, measured at 1K and B$\,$=$\,$14$\,$T. (a) Cut through $\nubg$$\,$=$\,$-1.  (b) Cut through $\nubg$$\,$=$\,$-2.  (c) Cut through $\nubg$$\,$=$\,$-3. (d) Cut through $\nubg$$\,$=$\,$-4. The conductance plateaus resulting from partial edge state equilibration at $\nutg$$\,$=$\,$-4  are highlighted in red and blue, assuming the 4th edge state is polarized (+,$\downarrow$) -Scenario I, or  (-,$\uparrow$) - Scenario II.}}
\end{figure*}

A map of $g(\Vtg,\Vbg)$ is shown at B$\,$=$\,$14$\,$T, T$\,=\,$1$\,$K on Figure 2(a). At this field, electron interactions are strong enough to fully lift the spin and valley degeneracy of the Landau levels~\cite{Zhang2006, Jiang2007, Goerbig2011, Kharitonov2012, Barlas2012, Alicea2007, Alicea2007bis, Sheng2007, Checkelsky2008, Nomura2006, Jung2009, Goerbig2006, Yang2006}.  As a result, when tuning $\nutg$ from -6 to 0 at constant $\nubg$$\,=\,$-6, $g$ shows plateaus at every multiple of $\Gq$ as fewer and fewer edge states are transmitted through the potential barrier [Fig. 2(b)]. These edge states can a priori be spin- and/or valley-polarized, depending on the ground-state in the bulk for each filling factor. It was pointed out in Ref.~\cite{Abanin2006} that the valley degeneracy of the zeroth Landau level is lifted by the confinement potential at the edges. As a result, the existence or lack of counter-propagating spin-polarized edge states at $\nu\,$=$\,0$ depends on the competition between Zeeman coupling and valley symmetry breaking interactions in the half-filled zeroth Landau level~\cite{Barlas2012, Jung2009, Abanin2007bis, Abanin2006}. For filling-factors $\nu\,$=$\,$1 and $\nu\,$=$\,$2, the edge states share a common valley polarization -- which may correspond to a superposition of the K and K$^{\prime}$ valleys, determined by the confinement potential~\cite{Abanin2006} -- but at $\nu$$\,=\,$2 the two edge states have opposite spins. As we move beyond $\nu\,$=$\,2$ we enter the $n\,$=$\,1$ Landau level. At quarter-filling ($\nu\,$=$\,3$), the valley polarization is arbitrary but Zeeman coupling favors an aligned spin polarization~\cite{Alicea2007bis, Sheng2007, Nomura2006}. At half-filling ($\nu\,$=$\,4$) electron correlations can favor either a spin or valley polarized ground-state~\cite{Young2012}, but not both. We label these Scenarios I and II respectively and show the associated two sequences of edge states in Fig. 2(c). In the following, we call the edge states' spin- and valley-polarization $\uparrow$/$\downarrow$ and +/- respectively. 

Fig. 3(a) plots $g$ vs. $\nutg$ for a fixed $\nubg$$\,$=$\,$-1.  At low density under the top gate, $g$$\,$=$\,$0, whereas around $\nutg$$\,$=$\,$$\nubg$$\,$=$\,$-1, $g$$\,$=$\,$1. From now on, we report conductance in units of $\Gq$ and define $g_{\nutg,\nubg}$=$g(\nutg,\nubg)$. For $\nutg$$\,$=$\,$-2 and -3, plateaus in conductance occur, although with less quality than for $\nutg$$\,$=$\,$-1. To quantify $g$ for $\nutg$$\,$=$\,$-2 and -3, we take an average over the plateau to obtain $g_{\rm{exp}}$$\,$=$\,$0.98$\pm$.04 and 0.66$\pm$.005, where the error is 1$\sigma$. A similar plot for $\nubg$$\,$=$\,$-2 is shown in Fig. 3(b).  Table I lists the expected values of $g$ for full equilibration of edge states ($g_{\rm{full}}$)~\cite{Abanin2007}, as well as the measured conductances $g_{\rm{exp}}$ for $\nutg$ up to 3. Deviations occur, for example $g_{2,1}\approx 1$ [Fig. 3(a)], instead of 2/3 as expected for full equilibration, indicating that edge states may not equilibrate.

\begin{table}[ht!]
\caption{Equilibration plateaus for $\nu_{B,T}\leq3$}
\vspace{5mm}
\begin{minipage}{.24\textwidth }%
\footnotesize
\begin{tabular}{ |c|c|c|c|c|c| } \hline
$\nubg$ & $\nutg$ & $g_{\rm{full}}$ & $g_{\rm{partial}}$ & $g_{\rm{exp}}$ & Edge state polarization\\ \hline
1 & 2 & 2/3 & 1 & 0.98$\pm$.04 & \hspace{32mm}\\ [5mm]
 1 & 3 & 3/5 & 2/3 & 0.66$\pm$.005 & \hspace{5mm}\\[5mm]\hline
\multirow{1}{*}{2}  & 3 & 1.5 & 5/3 & 1.68$\pm$.01 & \hspace{32mm} \\[5mm] \hline
\end{tabular}
\end{minipage}%
\begin{minipage}{.2\textwidth }%
\hspace*{0pt}\includegraphics[width=1\textwidth]{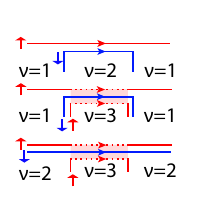}%
\end{minipage}
\end{table}

$g_{2,1}\approx 1$ suggests that the two opposite-spin edge states at $\nutg$$\,$=$\,$-2 do not equilibrate on the scale of our device, the $\uparrow$ edge state being fully transmitted through the $\nutg$$\,=\,$-2 region without scattering. The corresponding $g$=$\,$1 plateau extends at $\nubg$$\,$=$\,$-1 from $\nutg$$\,$=$\,$-1 to $\nutg$$\,$=$\,$-2 with a steep transition occurring only at $\nutg$$\,$=$\,$-3 [Fig. 2(a)]. This decrease in $g$ between $\nutg$$\,$=$\,$-2 and $\nutg$$\,$=$\,$-3 and the existence of well defined plateaus for $g_{3,1}$ and $g_{3,2}$ suggest that partial equilibration is due to the additional $\uparrow$ edge state at $\nu$$\,=\,$-3 --- despite its valley polarization, different from that of the zeroth Landau level's edge states. The plasma-etched edges of our samples are highly disordered on the lattice scale, which is expected to induce strong intervalley scattering. We therefore assume that the spin polarization is robust but that edge states with the same spin equilibrate regardless of their valley polarization [red region in Table I]. Consequently we propose a simple model for $g$ closely following Ref.~\cite{Abanin2007}, but where only edge states with identical spins equilibrate, regardless of their valley polarization~\cite{Suppinfo}. We use $\nu_{T,\uparrow}$($\nu_{T,\downarrow}$) to refer to the number of edge states with spin-polarization $\uparrow$($\downarrow$) at filling factor $\nutg$ under the top-gate, with similar notations for $\nubg$. The predicted conductance is then~\cite{Abanin2007}
\begin{equation}
g_{\rm{partial}}=\sum_{i=\downarrow,\uparrow}\frac{\nu_{T,i}\nu_{B,i}}{2\nu_{T,i}-\nu_{B,i}}
\end{equation}
At $\nutg$$\,$=$\,$-3, only the two edge states polarized spin down would then equilibrate, but not the third with opposite polarization [blue arrow in Table I] . This yields g$_{3{,}1}$$\,$=$\,$2/3 and g$_{3{,}2}$$\,$=$\,$5/3, in excellent agreement with our data, as highlighted in Table I and with dashed black lines on Fig. 3(a-b). 

\begin{table}[ht!]
\caption{Equilibration plateaus involving $\nu_{B,T}$$\,$=$\,$4}
\begin{tabular}{|c|c|c|c|c|c|}
\hline
$\nubg$ & $\nutg$ & $g_{\rm{full}}$ & $g_{\rm{partial}}$(I) & $g_{\rm{partial}}$(II) & $g_{\rm{exp}}$\\ \hline
1 & 4 & 4/7 & 3/5 & 2/3 & 0.60$\pm$.03\\ 2 & 4 & 4/3 & 8/5 & 4/3 & 1.59$\pm$.005\\
 3 & 4 & 12/5 & 5/2 & 8/3 & 2.50$\pm$.002  \\ 4 & 6 & 3 & 18/5 & 3 & 3.61 $\pm$.01 \\ \hline
\end{tabular}
\label{table:nonlin}
\end{table}

The polarization of the fourth edge state at $\nutg$$\,$=$\,$-4 is different for Scenarios I and II, allowing us to discriminate between them by measuring the conductance plateaus at $\nutg$$\,$=$\,$-4 for different values of $\nubg$. Using our simple model, we show in Table II the expected conductance plateaus $g_{\rm{partial}}(I)$ and $g_{\rm{partial}}(II)$ for the two Scenarios, respectively. In Scenario I, three $\downarrow$ polarized edge states equilibrate separately from the fourth edge state, which has opposite polarization. This differs from Scenario II where there are two edge states of each polarization, and each of those pairs equilibrates separately. For example, Equation (1) yields g$_{4{,}3}$$\,$=$\,$5/2 for Scenario I and 8/3 for Scenario II. Overall, our observed $g_{\rm{exp}}$ listed in Table II accords very well with Scenario I (spin-polarized half-filled n$\,$=$\,$1 level), in agreement with recent transport measurements~\cite{Young2012}. In Fig. 3(a-d), red and blue lines represent the expected plateaus at $\nutg\,=\,$-4 for Scenarios I and II, the match to data being noticeably better for Scenario I.  For example, the cut through $\nubg$$\,$=$\,$-1 in Fig. 3(a) would not show a transition at $\nutg$$\,$=$\,$-4 if the first and fourth edge states at $\nutg$$\,$=$\,$-4 did not have the same spin polarization.

In the bipolar regime, nonzero conductance implies disorder-mediated equilibration between edge states at the PN interface~\cite{Abanin2007,Long2008}. Full equilibration would lead to a quantization of the conductance $g$$\,$=$\,$$\nutg\nubg$/(2$\nutg{+}\nubg$), which we do not observe in any of our devices. Even at low field~\cite{Suppinfo}, $g$ is not quantized but remains smaller than $\Gq$, contrary to what has been observed  in more disordered samples. $g$ is strongly suppressed as B increases [Fig. 2], concomitant with the opening of the $\nu\,=\,0$ gap, and reaches on order 10$^{-7}\,$S at B$\,$=$\,$14$\,$T. We attribute this vanishing $g$ to the valley symmetry breaking of the zeroth level. At high fields, our data are consistent with a narrow insulating region ($\nu$$\,=\,$0 gapped state) spatially separating counter-propagating p and n edges, suppressing inter-edge state scattering~\cite{Suppinfo}. 

Similar local gating measurements on GaAs/AlGaAs two dimensional electron gas provided considerable insight into the physics of the integer and fractional quantum Hall effect, enabling for example interferometry~\cite{Chamon1997}, equilibration~\cite{Muller1991} and shot noise~\cite{Picciotto1997} measurements. We showed here that comparable studies in graphene and its multilayers are now within reach, with more complex phenomena likely to arise from the additional valley degree of freedom and the ambipolar band structure of graphene.


We thank D. Abanin for his input. This work was funded by the Center for Probing the Nanoscale, an NSF NSEC, supported under grant No. PHY-0830228. J. R. W. and D. G.-G. acknowledge funding from the W. M. Keck Foundation.

\clearpage

\section{Supplementary information}
\setcounter{figure}{0}
\renewcommand{\thefigure}{S\arabic{figure}}

\subsection{Device fabrication}

Polyvinyl alcohol (2$\%$ in water) is spun at 6000\,rpm on a bare silicon substrate and baked at 160\,$^{\circ}$C for 5\,min, resulting in a 40\,nm thick layer. Then, a layer of PMMA (950k, 5$\%$ in anisole) is spun at 2400\,rpm and baked 5\,mins at 160\,$^{\circ}$C. The total polymer thickness is approximately 450\,nm. Graphene is exfoliated onto this stack using Nitto tape (model 224LB), located using optical microscopy, and characterized with Raman spectroscopy. h-BN is exfoliated on a second silicon wafer piece with a 300nm thick thermal oxide, then heat-treated in Ar/O$_{2}$ at 500$^{\circ}$C for 8 hours to remove organic contamination [S1]. Only h-BN flakes that are atomically flat under AFM and show a pristine Raman spectrum are used for devices. 

The PVA layer is dissolved in deionized water at 90\,$^{\circ}$C, which lifts-off the PMMA membrane with the graphene attached to it. The membrane is then adhered across a hole in a glass slide~[S2] and baked at 110\,$^{\circ}$C. We then use a modified probe-station to align the slide with the graphene flake on top of the boron nitride substrate. Once both flakes are in contact, the stack is baked at 120\,$^{\circ}$C on a hot plate for 10 minutes to promote adhesion. The PMMA layer is dissolved in hot acetone, then rinsed in IPA, leaving the graphene flake on top of the boron nitride flake. This stack is annealed in flowing Ar/O$_{2}$ at 500\,$^{\circ}$C for 4 hours, which removes process residue and leaves the graphene flake pristine, as checked by Raman spectroscopy [S1]. 

We use e-beam lithography to pattern a graphene Hall bar (etched in oxygen plasma), then contact it with Cr/Au (1nm/100nm). In order to fabricate a suspended top gate above the device~[S3-6], the samples are spin-coated at 6000\,rpm with a solution of polymethyl-methacrylate (PMMA 950k), 3$\%$ in anisole, then baked at 160\,$^{\circ}$C for 5 minutes. An additional layer of methyl-methacrylate  (MMA 8.5$\%$ in ethyl lactate) is spin-coated at 6000\,rpm and baked 5 minutes at 160\,$^{\circ}$C. The MMA layer is 50$\%$ more sensitive to electron irradiation than the PMMA layer and it is therefore possible to develop the top resist layer without exposing the bottom layer.  The e-beam writing system we used is a JEOL 6300, with an acceleration voltage of 100\,keV. The contacts and the feet of the suspended bridge are exposed with 650\,$\mu \mathrm{C/cm}^{2}$, which is enough to dissolve both resist layers upon development in MIBK/IPA (1:3) for 45\,sec. The span of the suspended bridge is exposed with a base dose of \,290$\mu \mathrm{C/cm}^{2}$, which only develops the top resist layer. After development, the device is cleaned for 2 minutes with UV ozone, then metallized with 1nm of chromium and 150\,nm of gold. 

Some of our devices were current annealed in vacuum, flowing a current on the order of 1mA.$\mu m^{-1}$ for a few tens of seconds, which improved their mobility especially  on the hole-side.
 \begin{figure}
\center \label{figS1}
\includegraphics[width=3 in]{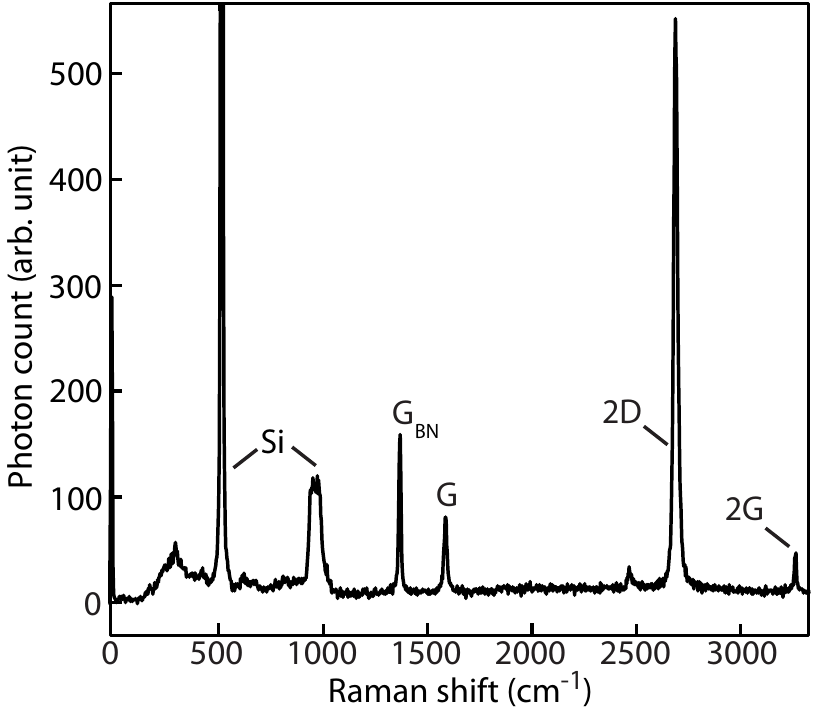}
\caption{\footnotesize{Raman spectrum of a monolayer graphene flake transferred on boron-nitride.}}
\end{figure}
\subsection{Raman spectra}
We use Raman spectroscopy to identify single-layer graphene flakes and estimate their degree of organic contamination. Figure S1 shows the Raman spectrum of a graphene flake transferred on boron nitride after Ar/O$_{2}$ annealing, but before the suspended-gate is fabricated. The ratio of the amplitudes of the 2D and G peak is $\mathrm{I}_{\mathrm{2D}}/\mathrm{I}_{\mathrm{G}}=7$, and the full width at half maximum of the 2D peak is 20 cm$^{-1}$, indicating that the flake studied here is monolayer graphene~[S7]. In the presence of organic contamination, the Raman spectrum would show a broad background signal in addition to the silicon, boron-nitride and graphene peaks. The absence of this background signal, the absence of a graphene D-peak (which would be visible on the left flank of the BN G peak), and the large 2D-to-G peak-ratio attest to the cleanliness of this device.

\begin{figure*}
\center \label{figS2}
\includegraphics[width=7 in]{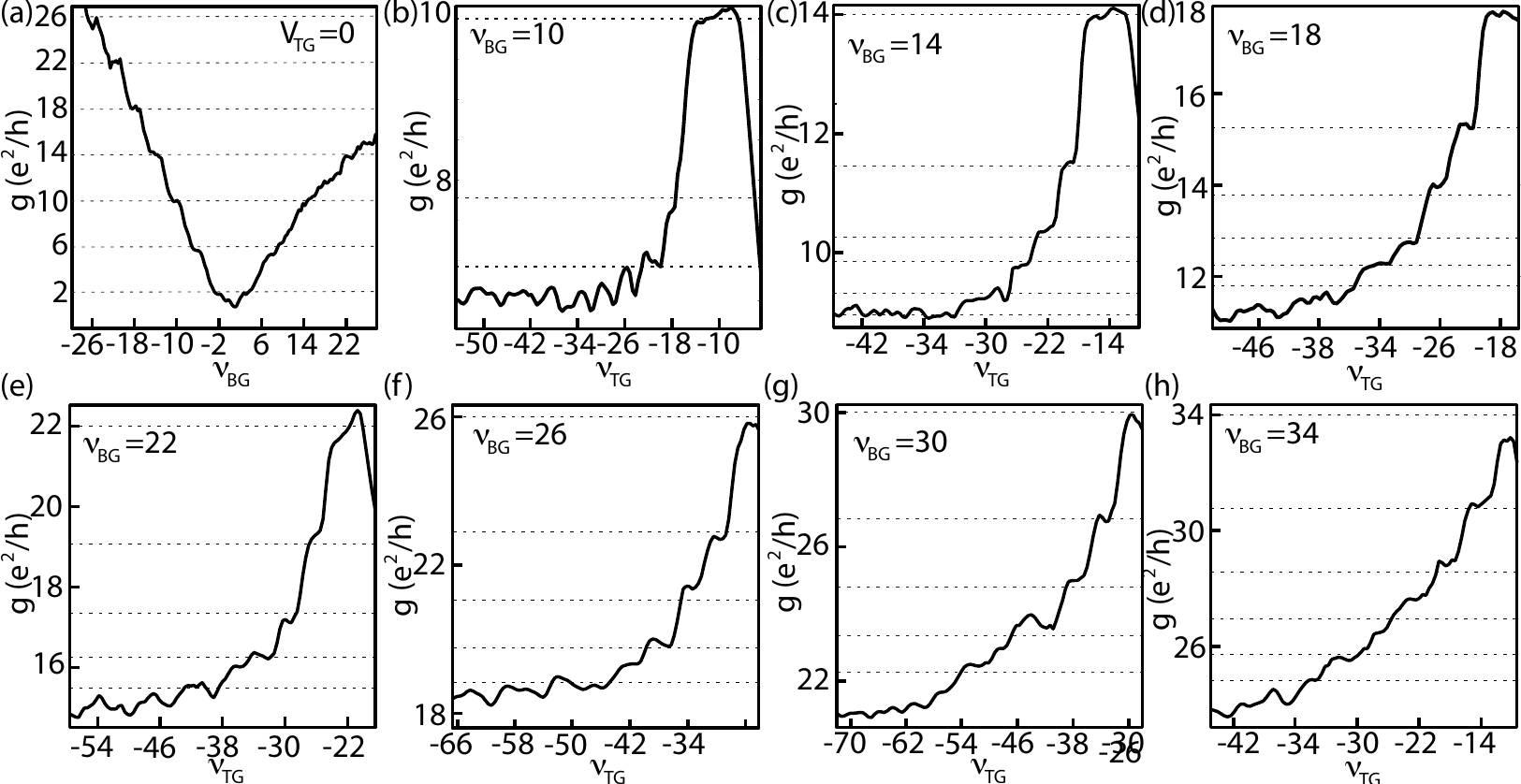}
\caption{\footnotesize{Cuts of the two-terminal conductance $g(\nutg)$ at constant $\nubg$ in the unipolar regime for filling factors $\nubg$$\,$=$\,$10 to 30, measured at 250mK and B$\,$=$\,$1T.}}
\end{figure*}

\begin{figure}
\center \label{figS3}
\includegraphics[width=3 in]{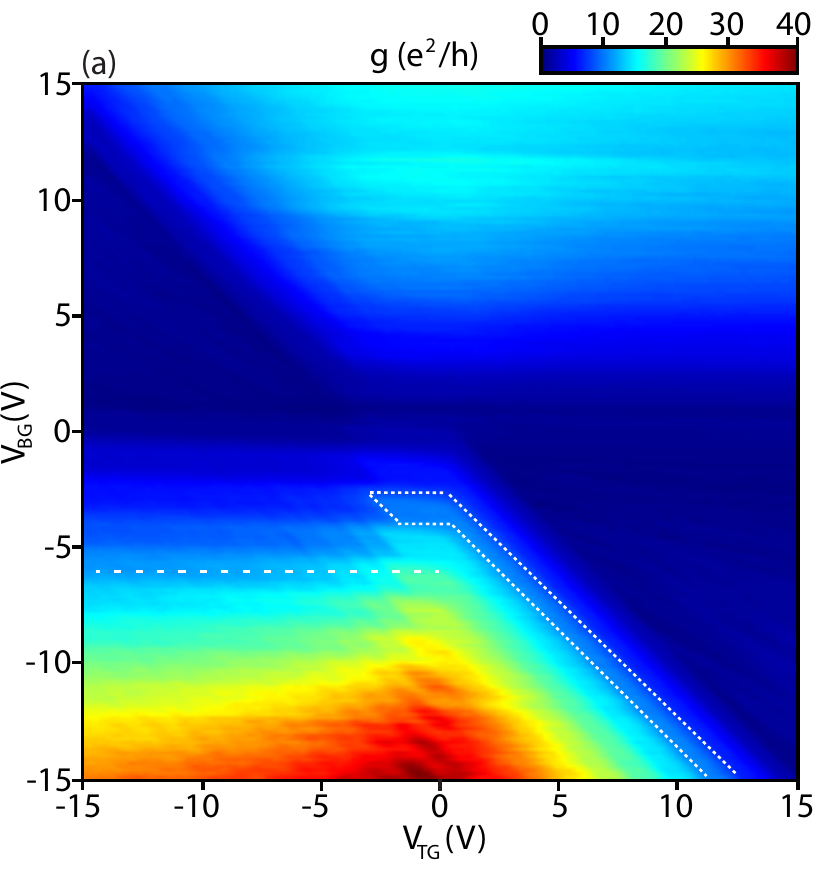}
\caption{\footnotesize{Two terminal conductance $g$, measured at B=1T and T$\,=\,$250mK as a function of the top- and back-gate voltages $\Vtg$ and $\Vbg$.}}
\end{figure}

\subsection{Conductance plateaus of top-gated devices with degenerate Landau levels}

%


We show on Figure S3 a map of $g(\Vbg,\Vtg)$ at B=1T, T=2K. This field is strong enough for our device to be in the integer quantum Hall regime, but Landau levels are still spin- and valley-degenerate. When the density is uniform across the flake, $g$ is well quantized on the hole-side and matches the expected sequence of plateaus for single-layer graphene: $g$$\,$=$\,$4(n+1/2), as highlighted with horizontal dashed lines in Fig. S2(a). The quantization is less visible on the electron side, which we found to be common in two-terminal graphene-on-BN devices. 

Scattering along the edges is strong enough to fully equilibrate degenerate edge states in the unipolar regime, in agreement with Ref.$\,$\cite{Abanin2007}. The expected conductance with full equilibration is $g_{\nutg,\nubg}\equiv g(\nutg,\nubg)$$\,$=$\,$$\nutg\nubg$/(2$\nutg{-}\nubg$). We show on Figure S2 cuts of the two-terminal conductance taken at constant filling-factor $\nubg$ outside the top-gate, we plotted dashed lines corresponding to the fractional plateaus expected from the above formula. As an example, Fig. S2(d) shows a cut of $g_{\nutg,18}$, corresponding to the dashed line on Fig. S3, with previously unobserved plateaus at $\nutg$$\,$=$\,$-18, -22, -26, -30, -34 and -38. The corresponding quantized values of $g$ are marked with dashes and respectively predicted to be: 18, 198/13, 234/17, 90/7, 306/25 and 342/29. The agreement with our data is excellent, which shows that degenerate edge states fully equilibrate in the unipolar regime, contrary to what is seen with polarized edge states at higher field. Most of these plateaus were not seen in previous dual-gated quantum-Hall-measurements, although the physical mechanisms involved here are similar to what was reported in Ref.~\cite{Williams2007, Ozyilmaz2007} of the main paper. 

Even at such low fields, we observe some discrepancies with previous experimental work~\cite{Williams2007, Ozyilmaz2007,Velasco2012}. While the quantized values of $g$ match the theoretical prediction for full mixing, we notice an anomalous shape of the plateaus in Fig. S3 (outlined in white dots), especially at lower $\nutg$, $\nubg$. At fixed $\nubg$, $g$ is expected to increase as $g_{\nutg,\nubg}$\,$=$\,$\vert\nutg\vert$ and start decreasing due to equilibration as soon as $\vert\nutg\vert>\vert\nubg\vert$, as shown in Fig. 2(b). Experimentally, however, equilibration only starts at $\vert\nutg\vert\ge\vert\nubg\vert+4$, giving the plateaus the shape of an elbow, as highlighted with dashes for the $g$$\,$=$\,$10 plateau around $\Vbg$$\,$=$\,$-3V and $\Vtg$$\,$=$\,$0V.  

\subsection{Equilibration and disorder}
\begin{figure}
\center \label{figS2}
\includegraphics[width=3 in]{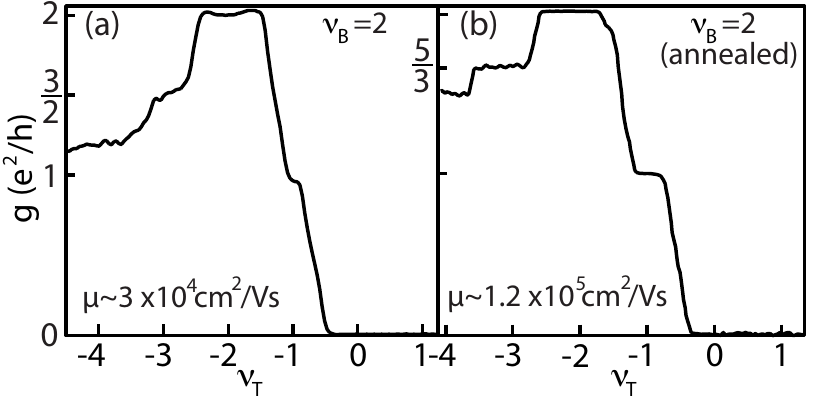}
\caption{\footnotesize{Cuts of the two-terminal conductance of Device B $g(\nutg)$ at $\nubg\,=\,-2$, measured at T=400mK and B=15T, before (a) and after (b) current-annealing.}}
\end{figure}

Figure S4 shows the two terminal conductance of a different device, Device B, measured at 400mK and B=15T. This device was initially of lesser quality than Device A and was measured before and after current-annealing in-situ. The carrier-mobility at low hole density is $\mu\,=\,$30000 cm$^{2}$/Vs before current annealing and 120000 cm$^{2}$/Vs after. Fig. S4(a) shows the cut at constant filling factor outside the gate $\nubg\,=\,$-2 as a function of $\nutg$ before annealing. The conductance plateau $g_{3,2}\,=\,1.48\pm.03$ is less pronounced but is close to the value 3/2 expected for full-mixing.  After annealing [Fig. S3(b)] the conductance is higher in the regime $\vert\nutg\vert>\vert\nubg\vert$, indicating partial equilibration. Plateaus are much flatter with $g_{3,2}\,=\,1.66\pm.01$ and $g_{4,2}\,=\,1.49\pm.01$, in very good agreement with the data from Device A. In particular $g_{3,2}$ is very close to the 5/3 plateau our model predicts.

\subsection{Map of the conductance for Device C}

Figure S5 shows the conductance $g(\Vtg,\Vbg)$ of another top-gated device, Device C, measured at 250mK and B=10T. The carrier mobility of this device is lower than Device A, on the order of 60000 cm$^{2}$/Vs at low temperature. $g$ is strongly suppressed below our noise floor around $\nubg$$\,$=$\,$0. Surprisingly, there is no corresponding continuous diagonal region with g$\,\approx\,$0 around $\nutg$$\,$=$\,$0, where the conductance should be suppressed under the gate. Instead, the diagonal edges of the regions of nonzero conductance (in the p-p$^{\prime}$-p and n-n$^{\prime}$-n regime respectively) are shifted inward, and the edges of the $g$$\,$=$\,$1 plateaus occurring at $\nutg$$\,$=$\,$$\pm 1$ are almost perfectly aligned as outlined by a white diagonal dashed-line spanning across Fig. 4. We observed this shift in all our devices, independent of the gate sweep speed and direction, ruling out hysteresis as the cause. Low quantum capacitance at $\nutg$$\,$=$\,$0 could broaden the $\nutg$$\,$=$\,$0 plateau, but, to our knowledge, should not introduce such an inward shift of conductance in the two quadrants. 

\vspace{2cm}
\begin{figure}[t!]
\center \label{figS2}
\includegraphics[width=3 in]{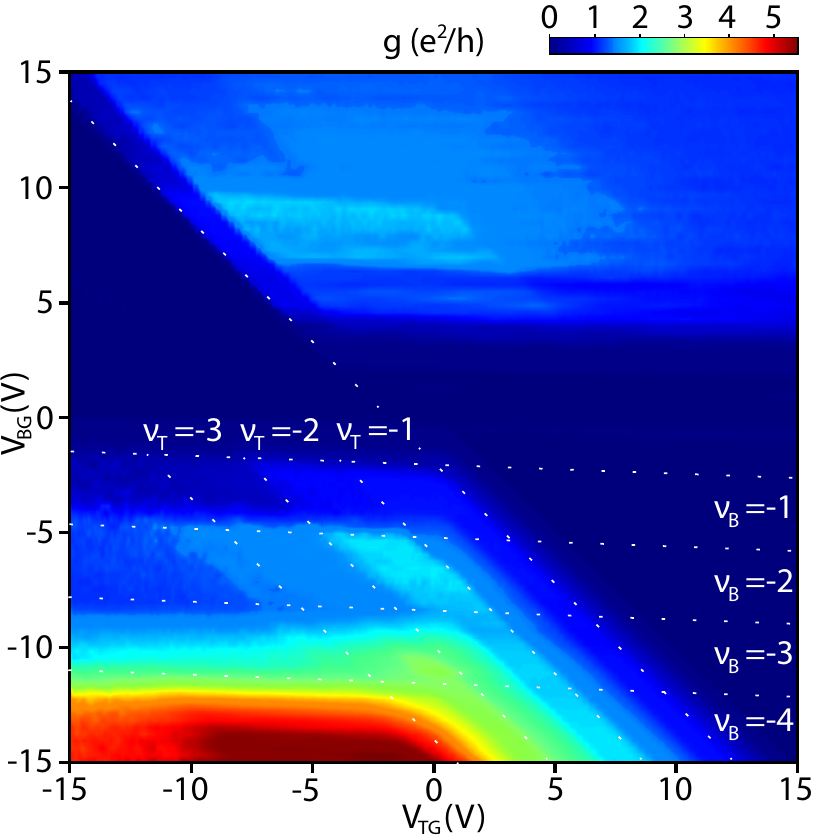}
\caption{\footnotesize{Two-terminal conductance $g$ of Device C as a function of both gate-voltages $\Vtg$ and $\Vbg$, measured at 250mK and B=10T.}}
\end{figure}

\section{Supporting references}

\begin{itemize}
\item[S1. ]
A. G. F. Garcia, M. Neumann, F. Amet, J. R. Williams, K. Watanabe, T. Taniguchi and D. Goldhaber-Gordon, Nano Lett. 12(9), 4449-4454 (2012).

\item[S2. ]
C. Dean, A. F. Young, I. Meric, C. Lee, L. Wang, S. Sorgenfrei, K. Watanabe, T. Taniguchi, P. Kim, K. L. Shepard and J. Hone, Nature Nano. \textbf{5}, 722Ð726 (2010).

\item[S3. ]
J. Velasco Jr, G. Liu, W. Bao and C. N. Lau, New J. Phys. \textbf{11}, 095008 (2009).

\item[S4. ]
R. T. Weitz, M. T. Allen, B. E. Feldman, J. Martin and A. Yacoby, Science \textbf{330}, 812 (2010).

\item[S5. ]
M. T. Allen, J. Martin and A. Yacoby, Nature Comm. \textbf{3}, 934 (2012).

\item[S6. ]
R. V. Gorbachev, A. S. Mayorov, A. K. Savchenko, D. W. Horsell and F. Guinea, Nano Lett. 8(7), 1995-1999 (2008).

\item[S7. ]
A. C. Ferrari, J. C. Meyer, V. Scardaci, C. Casiraghi, M. Lazzeri, F. Mauri, S. Piscanec, D. Jiang, K. S. Novoselov, S. Roth and A. K. Geim, Phys. Rev. Lett. \textbf{97}, 187401 (2006). 

\end{itemize}
\end{document}